\documentstyle[aps,prl,multicol,epsf]{revtex} 
\begin{document} 
\widetext
\title{\bf Metal-insulator crossover in the Boson-Fermion model 
in infinite dimensions}

\author{J.-M. Robin$^{(a)}$, A.~Romano$^{(b)}$ and J.~Ranninger$^{(c)}$}

\address{$^{(a)}$ Max Planck Institut f\"ur Physik komplexer Systeme,
Noethnitzerstrasse 38, D-01187 Dresden, Germany}

\address{$^{(b)}$ Dipartimento di Scienze Fisiche "E.R. Caianiello", 
Universit\`a di Salerno, I-84081 Baronissi (Salerno), Italy\\
Unit\`a I.N.F.M. di Salerno}

\address{$^{(c)}$ Centre de
Recherches sur les Tr\`es Basses Temp\'eratures, Laboratoire
Associ\'e \'a l'Universit\'e Joseph Fourier, 
\\ Centre National de la Recherche Scientifique, BP 166, 38042, 
Grenoble C\'edex 9, France}

\date{August 10, 1998} 
\maketitle 
\draft 
\begin{abstract}
The Boson-Fermion model, describing a mixture of tightly bound 
electron pairs and quasi-free electrons hybridized with each 
other via a charge exchange term, is studied in the limit of 
infinite dimensions using the Non-Crossing Approximation within 
the Dynamical Mean Field Theory. It is shown that a metal-insulator 
crossover, driven by strong pair fluctuations, takes place as the 
temperature is lowered. It manifests itself in the opening of a 
pseudogap in the electron density of states, accompanied by a 
corresponding effect in the optical and dc conductivity.
\end{abstract}


\begin{multicols}{2}

\narrowtext
The discovery of high-$T_c$ superconductors (HTS) has led to 
questioning a large number of basic physical concepts. 
In particular, great attention has been devoted to the
conjecture that in the normal state HTS do not behave
as standard Fermi liquids\cite{Anderson-97}. 
A signature for the possible breakdown of this concept might 
be the existence of a pseudogap in the normal state which 
opens up at a temperature $T^*$ which, depending on doping, 
can be well above the superconducting transition temperature 
$T_c$. The origin of this pseudogap is presently an issue 
of controversy. The fact that, upon lowering the temperature, 
this pseudogap smoothly merges into the superconducting gap 
in the entire Brillouin zone \cite{Ding-97}, strongly favors 
the idea that it is related to strong pair correlations which 
set in below a certain characteristic temperature $T^*$ 
and lead to condensation below $T_c$.
This does not mean that these pair correlations are due to some 
superconducting fluctuations but rather that they are a precursor 
to the superconducting state. They can exist without superconductivity 
and can in principle be stable all the way down to the lowest 
temperatures if the superconducting state is inhibited for some reason. 
If not, they will lead to condensation in some kind of superfluid state.
Local probe tunneling spectroscopy inside and outside vortex cores at 
temperatures below $T_c$ represents precisely this 
situation \cite{Renner-98}. Such experiments rule out a pseudogap due to 
superconducting fluctuations and moreover confirm that the normal 
state pseudogap goes over continuously into a true superconducting gap 
as one moves across the border of the vortex cores.

We have in the past attempted to understand the physics of HTS
in terms of a phenomenological model consisting of two components: 
localized pairs of electrons (Bosons) and quasi-free electrons 
(Fermions), hybridized with each other via a charge exchange term.
It is understood that the electrons are confined to the $CuO_2$ planes 
while the localized electron pairs are of the form of bipolarons and 
are confined to the dielectric highly polarizable layers which are 
sandwiching the $CuO_2$ planes\cite{Roehler-97}. The physics of 
this scenario is described by the Boson-Fermion model (BFM)
\begin{eqnarray}
H & = & \varepsilon_0\sum_{i,\sigma}c^{\dagger}_{i\sigma}c_{i\sigma}
-t\sum_{\langle i j\rangle,\sigma}c^{\dagger}_{i\sigma}c_{j\sigma}
\nonumber \\
& & + \; E_0 \sum_i b^{\dagger}_i b_i
+ g \sum_i [b^{\dagger}_i c_{i\downarrow} c_{i\uparrow}
+ c^{\dagger}_{i\uparrow} c^{\dagger}_{i\downarrow} b_i] \quad ,
\label{eq2}
\end{eqnarray}
with $\varepsilon_0=D-\mu$ and $E_0=\Delta_B-2\mu$. Here 
$c_{i\sigma}^{(\dagger)}$ denote fermionic operators for electrons with 
spin $\sigma$ at some effective sites $i$ (involving molecular units 
rather than individual atoms) and $b_i^{(\dagger)}$ are hard-core bosonic 
operators describing tightly bound electron pairs. The bare hopping 
integral for the electrons is given by $t$, the bare electronic
half bandwidth by $D$, the Boson energy level by 
$\Delta_B$ and the Boson-Fermion pair-exchange coupling 
constant by $g$. The chemical potential $\mu$ is common 
to Fermions and Bosons (up to a factor 2 for the Bosons) in 
order to guarantee charge conservation.

The underlying physics of this model is that the charge exchange term 
between Bosons and free electrons induces local pairing amongst the 
latter and thus leads to a depletion of the density of states near the 
Fermi level which reflects itself in the opening of a pseudogap. 
Our previous studies of this model on the basis of 
lowest order self-consistent conserving approximations 
bore this out\cite{Ranninger-95} and showed how the appearance of such a 
pseudogap manifests itself in the thermodynamic, 
transport and magnetic properties of the system\cite{Ranninger-96}.
These studies indicated that $T^*$ was intimately related to a precursor 
of a condensation of the intrinsically localized Bosons due to some 
highly non-linear feedback effect from the itinerant electrons. 
This finding was verified independently by a method summing up the 
most diverging diagrams\cite{Ren-98}.  
Our follow-up study of this model in the atomic limit\cite{Domanski-98} 
(which excluded from the outset any global superconducting state) gave 
the first indications that the opening of the pseudogap was 
related to a metal-insulator crossover rather than to superconducting
pair fluctuations. Both effects, of course, will determine the temperature 
dependence of the pseudogap, but with different importance depending 
on the temperature regime. We shall present in this Letter a dynamical 
mean field study of the BFM in infinite dimensions in which full use 
is made of those atomic limit results and which permits to incorporate 
the itinerancy of the electrons in a controlled fashion. We shall below 
sketch the main lines of this algorithm and then concentrate on the 
normal state properties of this model.

In infinite dimensions a many body system can be treated as a purely
local system coupled to a "medium" described by a Weiss field, with
which it interacts in the form of an impurity Anderson problem
\cite{Georges-96}. In such a scheme the effective Hamiltonian 
for the BFM is
\begin{eqnarray}
H & = & \sum_{\sigma} \varepsilon_{0} c_{\sigma}^{\dagger} c_{\sigma}
\; + \; E_{0} b^{\dagger} b \; + \; g \; [ \; c_{\uparrow}^{\dagger}
c_{\downarrow}^{\dagger} b \; 
+ \; b^{\dagger} c_{\downarrow} c_{\uparrow} \; ] \nonumber \\
  &   & + \; \sum_{k,\sigma} w_{k} \,
d_{k,\sigma}^{\dagger}d_{k,\sigma} \; + \; \sum_{k,\sigma} \; 
v_{k} \; [ \; d_{k,\sigma}^{\dagger}
c_{\sigma} \; + \; c_{\sigma}^{\dagger} d_{k,\sigma} \; ]
\end{eqnarray}
where $c_{\sigma}^{(\dagger)}$ and $b^{(\dagger)}$ denote the 
original Fermion and Boson operators at the impurity site, and
$d_{k,\sigma}^{(\dagger)}$ denote the auxiliary Fermionic variables of 
the Weiss field. For these latter the energy spectrum $w_{k}$ 
and their coupling $v_{k}$ to the impurity electrons have to be 
determined self-consistently. We see that the first line of Eq.(2) 
coincides with the local part of the Hamiltonian (1). 
Setting $H=H_0+H_I$, with $H_I$ denoting the interaction term between 
the electrons and the auxiliary Fermions,
we have that $H_0$ can be written in the form
$H_0= \sum_{n=1}^{8} \; | n \rangle \; E_{n} \; \langle n | \; 
+ H_{med}$, where $H_{med} = \sum_{k,\sigma} \; 
w_{k} \, d_{k,\sigma}^{\dagger}d_{k,\sigma}$ and 
\begin{equation}
\begin{array}{ll}
| 1 \rangle \; = \; | 0 \rangle & E_{1} \; = \; 0 \\
| 2 \; (3) \rangle \; = \; | \uparrow (\downarrow) \rangle 
& E_{2} \; = \; E_3 \; = \varepsilon_{0} \\
| 4 \rangle \; = \; u | \uparrow \downarrow \rangle \; 
- \; v | \bullet \rangle \qquad & E_{4} \; = \; \varepsilon_{0} + 
E_{0}/2 - \gamma \\
| 5 \rangle \; = \; v | \uparrow \downarrow \rangle \;
+ \; u | \bullet \rangle & E_{5} \; = \; \varepsilon_{0} + 
E_{0}/2 + \gamma \\
| 6 \; (7)\rangle \; = \; | \uparrow (\downarrow) \;
\bullet \rangle & E_{6} \; = \; E_7 \; = \varepsilon_{0} + E_{0} \\
| 8 \rangle \; = \; | \uparrow \downarrow \bullet \rangle &
E_{8} \; = \; 2 \varepsilon_{0} + E_{0}
\end{array}
\label{eigen}
\end{equation}
are the eigenstates and the eigenvalues of the Hamiltonian (1) in the
atomic limit. Here the arrows denote the presence of an electron with 
spin up or down and the dot indicates the presence of a Boson on 
the effective impurity site. In Eqs.(\ref{eigen}) we have defined
$u^{2}=1/2-(2\varepsilon_{0} - E_{0})/(4\gamma)$, $v^{2}=1-u^2$ and 
$\gamma=\sqrt{(\varepsilon_{0} - E_{0}/2)^{2} + g^2}$.

Introducing the Hubbard operators $X_{mn} = | m \rangle \langle n |$
we can rewrite the original Fermion operators as
\begin{eqnarray}
c_{\sigma} \; = \; \sum_{mn} \; F_{mn}^{\sigma} \; X_{mn}
\; \; , \; \; \; 
c_{\sigma}^{\dagger} \; = \; \sum_{mn} \; 
F_{nm}^{\sigma} \; X_{mn}
\end{eqnarray}
where
\begin{eqnarray*}
F_{12}^{\uparrow} &=& F_{78}^{\uparrow} = 1, \; 
F_{34}^{\uparrow} = F_{56}^{\uparrow} = u, \;
F_{35}^{\uparrow} = - F_{46}^{\uparrow} = v \nonumber \\
F_{13}^{\downarrow} &=& - F_{68}^{\downarrow} = 1, \; 
F_{24}^{\downarrow} = - F_{57}^{\downarrow} = - u, \; 
F_{25}^{\downarrow} = F_{47}^{\downarrow} = - v 
\end{eqnarray*}
The interaction Hamiltonian then becomes
\begin{equation}
H_{I}  =  \sum_{k \sigma} \sum_{mn} \, v_{k} \, [ \,
F_{mn}^{\sigma} \, d_{k,\sigma}^{\dagger}
X_{mn} \, + \, F_{nm}^{\sigma} \, X_{mn} d_{k,\sigma} \, ] \; .
\end{equation}
We shall now solve the $d\to\infty$ BFM within the so-called 
Non-Crossing Approximation (NCA) in $H_I$.
A similar approach was recently used in the Dynamical Mean Field 
Theory (DMFT) context\cite{Georges-96} by Lombardo 
{\it et al.} for a multiband Hubbard model for 
perovskites \cite{Avignon} and by Schork and Blawid for the Anderson 
lattice model with correlated conduction electrons \cite{Schork}.
For this purpose we introduce the resolvents $R_{mn}(z)$, where
$m$ can be different from $n$, defined through the Dyson equation
\begin{equation}
R_{mn}(z) = \delta_{mn}R^0_m(z) + 
\sum_{l} R^0_m(z)\,\Sigma_{ml}(z)\,R_{ln}(z) \; ,
\label{resol}
\end{equation}
together with the associated spectral functions given by
$A_{mn}(\omega)=-2$\,Im\,$R_{mn}(\omega + i\delta)$.
The free diagonal resolvents in Eq.(\ref{resol}) are 
$R^0_m(z)=(z-E_m)^{-1}$ and the self-energies 
$\Sigma_{ml}(z)$ are determined self-consistently through the equations
\begin{eqnarray}
&&\Sigma_{mm'}(z) = \sum_{nn',\sigma} 
F_{nm}^{\sigma} F_{n'm'}^{\sigma} \int \frac{dx}{2\pi} \, \Delta(x) \, 
n_{F}(-x) R_{nn'}(z-x) \nonumber \\
&& \quad \; + \sum_{nn',\sigma} 
F_{mn}^{\sigma} F_{m'n'}^{\sigma} \int \frac{dx}{2\pi} \; \Delta(x) \; 
n_{F}(x) R_{nn'}(z+x) 
\label{self}
\end{eqnarray}
with the use of the Weiss field spectral function 
\begin{equation}
\Delta(\omega) \; = \; 2 \pi \; \sum_{k} \; v_{k}^{2} \; \delta(\omega - 
w_{k}) \quad . 
\end{equation}
Within the NCA the local Fermion Green's function
\begin{equation}
G_F(z) = \sum_{mn} \sum_{m'n'} F^{\sigma}_{mn} F^{\sigma}_{n'm'} 
\langle\langle X_{mn} \, ; \, X_{m'n'} \rangle\rangle_{z}  
\end{equation}
is expressed in terms of the resolvents according to the relations
\begin{eqnarray}
\langle\langle X_{mn} \, ; \, X_{m'n'} \rangle\rangle_{z} & = &
{1\over Z_{loc}} \int \frac{dx}{2\pi} e^{-\beta x} 
[ A_{m'm}(x) R_{nn'}(x+z) \nonumber \\
& & + \; \xi A_{nn'}(x) R_{m'm}(x-z) ] \quad ,
\label{aver}
\end{eqnarray}
where  
\begin{equation}
Z_{loc} = \sum_{mm'} \int dx e^{-\beta x} A_{mm'}(x)
\end{equation}
and $\xi=-1$ for Fermions and $+1$ for Bosons.
The above set of self-consistent equations is evaluated for a Bethe 
lattice with a semi-circular density of states $D(\varepsilon)$ 
of width $2D=4t$. In this case the DMFT self-consistency condition 
takes the simple form $\Delta(\omega)\,=\,t^2\,A_F(\omega)$
\cite{Georges-96}, 
where $A_F(\omega)=-2\,$Im$\,G_F(\omega+i\delta)$ is the 
fermionic spectral function. 
We should stress that our approach represents a
generalization of the NCA (originally devised for the 
single-impurity Anderson model \cite{Bickers-87}), suited for 
models with a total number of particles per site greater than one. 
The original NCA formulation is recovered when all the non-diagonal 
resolvents are neglected, as was recently done for the multiband 
correlated electron models studied in Refs.\cite{Avignon,Schork}. 
In general, the iterative procedure determining those resolvents will 
lead to an increasing number of such non-diagonal elements, 
the relevance of which will depend on the model under consideration.
For the problem studied here, the only non-vanishing non-diagonal 
element is $R_{45}\,(=R_{54})$.

The results presented below have been obtained in the symmetric limit 
of the model, i.e., for $\varepsilon_0=E_0=0$ (from now on all energies 
are in units of $2D$). In this case the chemical 
potential is fixed in the middle of the electronic band, giving
$n_F=\sum_{\sigma }\langle c_{\sigma}^{\dagger}c_{\sigma} \rangle=1$ and 
$n_B=\langle b^{\dagger} b \rangle = \frac{1}{2}$. We furthermore 
choose $g=0.2$ such that the characteristic temperature of the opening 
of the pseudogap lies in the regime of a few hundred degrees $K$. 
From the behavior of the density of states $A_F(\omega)$, 
plotted in Fig.\,1 for various temperatures $T$, we can see that
the pseudogap opens up below a characteristic temperature 
$T^* \simeq 0.15$ and gradually deepens as $T$ is reduced.
This value of $T^*$ gives a reasonable estimate of the onset of the
pseudogap regime, provided that the Fermions are assumed to be 
quasi-particles with a strongly renormalized bandwidth.
We also notice that temperatures lower than those reported in the
figure are not considered here, since 
they do not lead to reliable convergency rates. This is probably a 
sign of the failure of the NCA at low temperatures.

The behavior of the pseudogap is best studied in terms of the 
fermionic self-energy $\Sigma_F(z)$ or its imaginary part 
$\Gamma(\omega) = - 2\,$Im$\,\Sigma_F(z = \omega + i\delta)$. 
$\Sigma_F(z)$ is introduced by expressing the local Green function 
in the form $G_{F}(z) \, = \, [z - \varepsilon_{0} - \Sigma^{W}(z) - 
\Sigma_F(z)]^{-1}$,
where the Weiss self-energy $\Sigma^{W}(z)$ is determined from
$\Delta(\omega) = - 2\,$Im$\,\Sigma^{W}(\omega + i\delta)$  
or, alternatively, using the self-consistency condition
$\Sigma^{W}(z) = t^{2} G_{F}(z)$. 
For an ordinary Fermi Liquid, one expects that 
$\Gamma(\omega) \sim \omega^{2}$ such that the lifetime of the 
quasi-particles at the Fermi 
energy $\tau_{F} = 1 / \Gamma(0)$ becomes infinite. In Fig.\,2a, 
we plot $\Gamma(\omega)$ for the same parameters as in Fig.\,1 
and observe that the pseudogap is linked to a strong resonant
scattering of the quasi-particles with the Bosons at the Fermi energy. 
As one lowers the temperature, the lifetime at the Fermi energy 
tends to zero and a metal-insulator crossover takes place. 
The form of $\Sigma_F$ clearly shows that our system
does not correspond to a Fermi liquid, as we can see from
an inspection of the poles of the lattice Green's 
function $G_F(z,\omega)=[z-\varepsilon_k-\Sigma_F(z)]^{-1}$ 
($\varepsilon_k$ denoting the bare electronic dispersion). At high 
temperatures, $T \gg T^*$, we see from Fig.\,2b that the straight line 
representing the 
\begin{figure}
\centerline{\epsfxsize=7cm \epsfbox{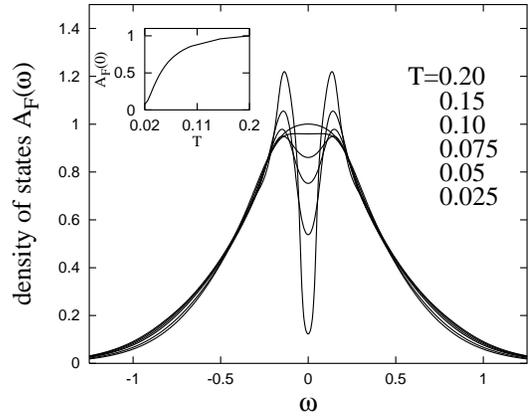}}
\caption{Fermion DOS for several temperatures $T$ (all energies in units 
of $2D$), the deepest pseudogap corresponding to the lowest $T$. The 
inset shows the DOS at the Fermi energy as a function of $T$.}
\label{f1dos}
\end{figure}
\noindent
inverse of the free-particle Green's function 
at $k=k_F$ (with $\varepsilon_{k_F}=0$) cuts Re$\,\Sigma_F(\omega)$ only
at the Fermi energy $\omega=0$. Considering, however, that $\Gamma(\omega)$ 
develops a minimum at $\omega=0$ which tends to a constant as  
$T$ is increased, we still cannot attribute this feature to a normal
Fermi liquid behavior. Upon lowering the temperature, we observe an 
inversion of slope which results in a cut occurring at three distinct 
frequencies, with the solution at $\omega=0$ being now accompanied by 
a large imaginary part of $\Sigma_F$. The pseudogap is thus a consequence
of a spectral weight transfer from the Fermi energy away to the wings 
in the density of states associated with bonding and anti-bonding 
two-particle states\cite{Domanski-98}. This is very 
similar to what has been obtained for the $2d$ Hubbard model at half 
filling\cite{Bumsoo-98}, where a pseudogap develops due to 
antiferromagnetic correlations. The results presented here 
are also in line with the precursor pairing correlation ideas developed by 
Randeria\cite{Randeria-98}.
\begin{figure}
\centerline{\epsfxsize=7cm \epsfbox{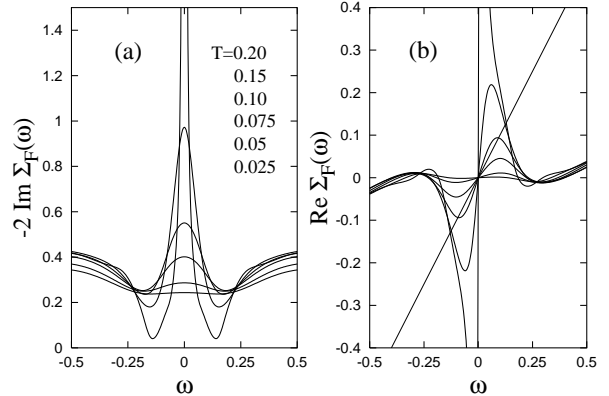}}
\caption{Frequency dependence of the imaginary (a) and the real (b) part 
of the Fermion self-energy for several temperatures $T$ (same units as 
in Fig.\,1). The lowest $T$ corresponds to the highest value of 
Im\,$\Sigma_F(0)$ and the steepest slope of Re\,$\Sigma_F(0)$,
respectively.}
\label{f2self}
\end{figure}

The present study does not contain any superconducting fluctuations and 
hence excludes the possibility of a transition to a superconducting 
state. The opening of the pseudogap is hence entirely related to a 
crossover from a metallic-like state at high temperatures to an insulating 
one at lower temperatures. We now show that this behavior is also 
manifest in the real part of the optical conductivity which in the 
limit $d\to\infty$ takes the form\cite{Georges-96}
\begin{eqnarray}
\label{opt}
\sigma(\omega)  & = & \pi \int\! d\varepsilon D(\varepsilon)\int\! 
d\omega ' \, 
A_F(\varepsilon,\omega')A_F(\varepsilon,\omega+\omega') \,\times
\nonumber \\
& & \times \; {1\over\omega} \left[n_F(\omega')-n_F(\omega+\omega')\right] 
\quad .
\end{eqnarray}
Here the sum over momenta in the spectral function 
$A_F(\varepsilon_k,\omega)$ of the lattice 
Green's function has been expressed as an energy integration over 
$D(\varepsilon)$. From the results presented in Fig.\,3, we notice 
that the opening and gradual deepening of the pseudogap, as the 
temperature is reduced, is associated with a change-over in the dc 
resistivity $\sigma(0)^{-1}$ from a linear in $T$ decrease to an 
upturn (see the inset). We stress that the linear behavior does not 
arise from the temperature dependence of the single-particle lifetime 
$\Gamma(0)^{-1}$, which for $T \gg T^*$ becomes independent on $T$, 
but is exclusively due to an intrinsic non-Fermi liquid behavior even 
above $T^*$. This is also manifest in the fact that the Fermion DOS 
never recovers the unperturbed semicircular shape, however high $T$ is.
As far as the frequency dependence of $\sigma(\omega)$  
is concerned, we observe below $T^*$ a well defined 
isosbestic point around which the spectral weight is shifted from 
low to higher frequencies and at which the various curves for the 
optical conductivity, corresponding to different $T$, cross.
This behavior, recently confirmed by optical measurements\cite{Timusk-97},
is in line with our previous results\cite{Ranninger-96} on 
the optical conductivity in the 1$d$ and 2$d$ BFM.

In conclusion, we have reported in this Letter on a metal-insulator 
crossover described by the BFM for HTS, driven by strong pair 
fluctuations and leading up to the opening of a pseudogap in the 
density of states. Our calculations are based on a generalization of 
the non-crossing approximation, within a dynamical mean field 
study in the $d \rightarrow \infty$ limit. The results obtained 
are similar to those for strongly correlated systems were 
the pseudogap is due to antiferromagnetic correlations.
In order to decide which of the mechanisms is relevant for HTS, a 
detailed knowledge is required of the $k$ dependence of this pseudogap 
in the Brillouin zone, together with the temperature behavior of 
the incoherent part of the single-particle spectrum. A study for the 
latter, on the basis of a generalized BFM in the atomic limit, 
associates it with phonon shake-off effects, when the Bosons are 
considered as bipolarons\cite{Ranninger-98}. Whatever the mechanism 
of pairing might be, the present study
of the BFM clearly suggests a normal state pseudogap due to 
uncorrelated pair fluctuations which, upon approaching $T_c$ from above, 
will tend to correlate and lead to a true superconducting gap.  
\begin{figure}
\centerline{\epsfxsize=7cm \epsfbox{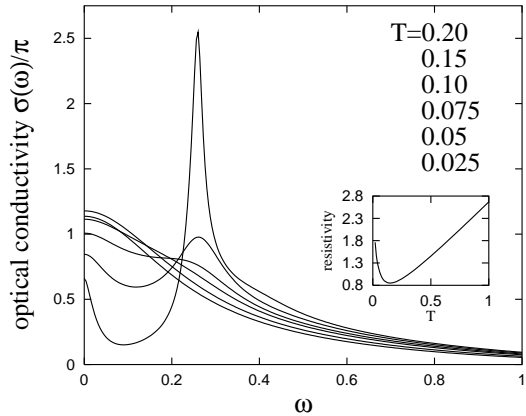}}
\caption{Fermion optical conductivity for several temperatures $T$, the 
lowest $T$ corresponding to the highest value of $\sigma(\omega)$ at
high frequencies (same units as in Fig.\,1). The inset illustrates the 
dc resistivity $(\sigma(0)/\pi)^{-1}$ as a function of $T$.}
\label{f3opt}
\end{figure}

\end{multicols}

\end{document}